\documentclass[referee]{aa} 
\usepackage{graphicx}
\usepackage{txfonts}
\begin{document}
   \title{C stars in the outer spheroid of  NGC 6822\thanks
{This publication makes use of data products from the Two Micron All Sky Survey, 
which is a joint project of the University of Massachusetts and the Infrared 
Processing and Analysis Center/California Institute of Technology, funded by 
the National Aeronautics and Space Administration and the National Science Foundation.}}
%


   \author{S. Demers
          \inst{1}
\and
           P. Battinelli \inst{2}
\and
	E. Artigau \inst{1}
         }

  \institute{ D\'epartement de Physique, Universit\'e de Montr\'eal,
                C.P.6128, Succursale Centre-Ville, Montr\'eal,
                Qc, H3C 3J7, Canada\\
                \email {demers@astro.umontreal.ca }
		\email{artigau@astro.umontreal.ca}
\and
   	     INAF, Osservatorio Astronomico di Roma
              Viale del Parco Mellini 84, I-00136 Roma, Italia\\
              \email {battinel@oarhp1.rm.astro.it }}

   \date{Received; accepted}


   \abstract
{From a 2$^\circ\times$ 2$^\circ$ survey of NGC 6822  we have previously
 established that this
Local Group dwarf irregular galaxy  possesses a huge spheroid
having more than one degree in length. This spheroid is in rotation
but its rotation curve is known only within $\sim 15'$ from the center.
It is therefore critical to identify bright stars belonging to the spheroid 
to characterize,
as far as possible, its outer kinematics.}
{We use the new wide field near infrared imager CPAPIR, operated by the SMARTS
consortium,  to acquire
$J$, $K_s$ images of two 34.8$'\times 34.8'$ areas in the outer spheroid to
search for C stars. 
}
{The colour diagram of the fields allows the identification of  192 
C stars candidates but a study of the FWHM of the images permits the 
rejection of numerous non-stellar objects with colours similar to C stars.
}
{ We are left with 75 new C stars, their mean K$_s$ magnitude and mean
colour are similar to the bulk of known NGC 6822 C stars.
}
{This outer spheroid survey confirms that the intermediate-age AGB stars are
a major contributor to the stellar populations of the spheroid. The discovery
of some 50 C stars well beyond the limit of the previously known rotation curve 
calls for a promising spectroscopic follow-up to a major axis distance of 40$'$. 
}

\keywords{ galaxies, individual: NGC 6822 -- galaxies: local group
-- stars: carbon
}


\maketitle
%

\section{Introduction}

NGC 6822, a Local Group dwarf irregular galaxy, has been the first
extragalactic object identified by Hubble (1925) who determined its
distance to be 214,000 pc or (m--M) = 21.65.  NGC 6822
is located in the constellation Sagittarius at a low Galactic latitude
($\ell$ = 25$^\circ$, b = --18$^\circ$). It is thus
seen behind a relatively heavy stellar foreground and its reddening 
is far from negligible. Today, we adopt for NGC 6822 a distance
of (m -- M)$_0$ = 23.35, based on the RR Lyrae observations of Clementini
et al. (2003), the Cepheids observations of Pietrzy\'nski et al. (2004)
and the near infrared observation of the tip of the red giant
branch (TRGB) by Cioni \& Habing (2005). At 470 kpc, it is the nearest
Magellanic-type galaxy, after the Magellanic Clouds.

Contrasting strongly with its discovery description by Barnard (1885)
who depicted it as: ``a 2$'$ nebula, diffuse and difficult to see
with his 5-inch instrument'', NGC 6822 has been found to be much more
extended. Its global structure was first studied 
by Hodge (1977) who described the bar  as a 8$'$ long structure 
with a position angle of 10$^\circ$. In a more
detailed investigation,  by Hodge et al. (1991), the galaxy could be traced
to 10$'$. They adopt the point of view that the galaxy is circular
rather than following the shape of the bar. We realize today that
the low-density of the
periphery of the galaxy can easily be masked by the substantial
foreground density seen along the line of sight. 

The first use of a wide-field imager, to search for C stars, by
Letarte et al. (2002) has revealed that NGC 6822 is surrounded by
a huge stellar
structure extending to at least 20$'$  from its center and containing a
substantial intermediate-age population. This huge spheroid was also seen
by Lee \& Hwang (2005).  Young stars were
identified far from the center, along the HI disk by Komiyama et al.
(2003). Recently, Battinelli et al. (2006) have mapped the elliptical 
spheroid of NGC 6822 to a semi-major axis distance of 36$'$ . From
radial velocities of numerous C stars, Demers et al. (2006) have shown that
the spheroid is dynamically decoupled from the HI disk and
rotates at right angles to it. Since C stars
used as kinematical probes were then known only to a distance of $\sim 15'$
, it is of great interest to identify other C stars further out
in the spheroid to extend its rotation curve  to compare it to the HI
rotation curve, observed to $\sim30'$ (Weldrake et al. 2003).

Few near-infrared
investigators have targeted NGC 6822. Elias \& Frogel (1985) have 
obtained $J$,$K$ magnitudes of 18 mostly spectroscopically confirmed 
red supergiants, with a $\langle K\rangle$ = 13. These stars are much
brighter than C stars. Davidge (2003) has observed three tiny areas
near the NGC 6822 bar to determine the metallicity from the slope
of the RGB. He concluded that the bulk of the NGC 6822 red giants are of
intermediate-age rather than $\sim 12$ Gyr.
The wide-field survey of a 20$'$ $\times$ 20$'$
central region of NGC 6822 by Cioni \& Habing (2005)
is a first attempt to present a global view of the AGB of NGC 6822 in the 
near infrared. More recently, Kang et al. (2006) have obtained $J,H,K$ photometry
of small areas near the center, where the space density of C stars is quite
high.

We then present results of a near-infrared survey, which targets the 
outer spheroid and takes
advantage of the new wide-field imager CPAPIR. It is essentially
complementary to Cioni \& Habing (2005) observations. We survey two 
regions in the outer spheroid of NGC 6822 with a small overlap with their
central field.

\section{Observations}

The  $J$, $K_s$ observations described here were obtained in service mode
(from April to September 2005)
with the CPAPIR ({\it Cam\'era Panoramique Proche InfraRouge,} see Artigau et al., 2004) 
imager attached to the CTIO 1.5 m telescope and operated by the SMARTS consortium.
CPAPIR is based on a 2048 $\times$ 2048  Hawaii-2 infrared
array detector. With a pixel size of 1.02$''$, at the 1.5 m telescope it
has a field of view of 34.2$'$ $\times$ 34.2$'$. 
One frame consists of five co-added exposures of 5.4 s for $K_s$ and a 
single 24.3 s for
$J$. The script, for each observation, acquires 100 frames  while 
introducing a 5$''$  dither after each frame. 
The sky image is built up by running a median
of nine frames, centered on the frame of interest.  Stars are masked 
during the sky construction.
These 100 frames are registered and a median image is built. 

The combined median images are then analysed with DAOPHOT-II (Stetson 1994). 
The instrumental magnitudes and colours are transformed into the
$J$, $K_s$ system by cross-identifying our program stars with the 2MASS
stars in the fields.  Nearly one thousand 2MASS stars
are seen in each field. Least square solutions, omitting the 
saturated bright stars and the faint 2MASS stars with poorer photometry,
yield:
$$ {\rm NE}\ \  J = -3.686 (\pm 0.046) + 1.005 (\pm 0.003)j_{inst}$$
$$ {\rm NE}\ \  K_s = -3.972 (\pm 0.069) + 0.993(\pm 0.004)k_{inst}$$
$$ {\rm SW}\ \  J = -3.703 (\pm 0.041) + 1.0064(\pm 0.002)j_{inst}$$
$$ {\rm SW}\ \  K_s = -4.262 (\pm 0.070) + 1.002(\pm 0.004) k_{inst}$$

These equations show that, as expected, the slopes are 1.00, we need only
to apply the magnitude zero point shifts. 

Figure 1 shows a SDSS image of 
NGC 6822 along with the outer ellipse (semi-major axis of 36$'$), 
defining the spheroid,
identified by Battinelli et al. (2006). The two observed CPAPIR fields, 
we called NE and SW, 
are outlined. Their coordinates are given in Table 1 along with the
exposure times, in seconds, for each filter.
   \begin{table}
      \caption[]{NGC 6822 observations with exposure times}
    $$
       \begin{array}{lccc}
            \hline
            \noalign{\smallskip}
            {\rm Field}&{\rm RA\ J2000\ Dec}&{\rm J}&{\rm K_s}  \\
           \noalign{\smallskip}
            \hline
            \noalign{\smallskip}

{\rm NE}&19:46:34.1\ -14:38:21&6846\ s&10017\ s\\
{\rm SW}&19:43:25.8\ -14:58:38&6092\ s&7304\  s\\
            \noalign{\smallskip}
            \hline
         \end{array}
     $$
   \end{table}
   \begin{figure*}
   \centering
\includegraphics[width=9cm]{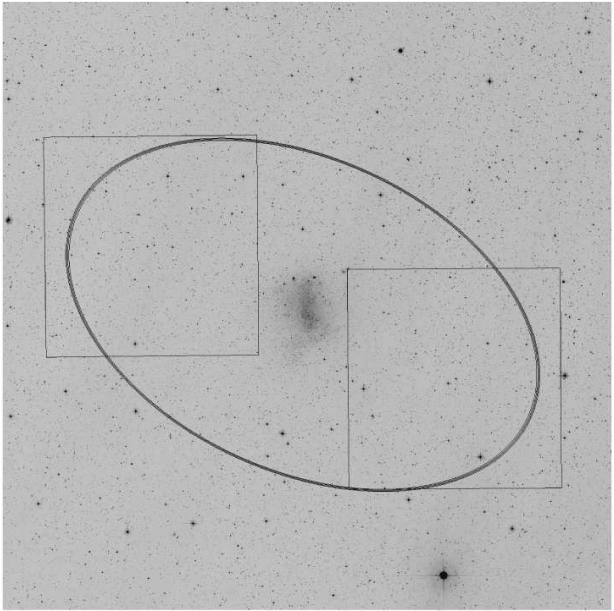}

   \caption{NGC 6822, the outer 36$'$ ellipse is sketched along with
the two CPAPIR fields, NE at the upper left and SW at the lower right,
 discussed in this paper. 
}
              \label{ Fig field}
    \end{figure*}

\section{Results}
The colour-magnitude diagram of the combined two fields
 is presented in Figure 2.
Some 18,000 stars with colour errors
$\sigma_{jk}$ $<$ 0.20 are plotted. A global reddening of E(B--V) = 0.25 
is assumed (van den Bergh 2000; Battinelli et al. 2006). Using the 
Schlegel et al. (1998) relations, this reddening translates into
$E(J-K)$ = 0.13 and A$_K$ = 0.09, small but nevertheless not negligible values. 
The vertical line, drawn on the CMD at $(J-K_s)$ = 1.49 corresponds to
the adopted separation of O-rich and C-rich AGB stars, $(J-K_s)_0$ = 1.36 
by Cioni \& Habing (2005). The horizontal dashed line is the level of the TRGB,
at $K_s$ = 17.1 (Cioni \& Habing 2005).
Our 360 stars in common with their photometry,
(for which their $\sigma_k < 0.20$) yield: 
$\Delta_{K_s}$ = --0.06 $\pm$ 0.32, while our  185 stars (for which their
$\sigma_{jk} < 0.30$) yields:  $\Delta_{J-K}$ = --0.05 $\pm$ 0.24.
Our observed regions do not overlap with the near infrared photometry
recently published by Kang et al.  (2006).

Our CMD should be compared to Cioni \& Habing (2005)
 Fig. 7 and Fig. 10. However, our CMD contains  a much
larger contribution from the foreground/background because the areas observed 
are more than twice as large as their region and, most importantly because
we target the outer parts of the galaxy
where the proportion of NGC 6822 stars is low. The two major vertical 
features seen on the CMD are Galactic G dwarfs, on the left and the
Galactic M dwarfs at $(J-K_s) \approx$ 1.0. Actually, since our photometry
reaches barely magnitudes fainter than the tip of the red giant branch, 
the minority of the stars seen here belongs to NGC 6822.

   \begin{figure*}
   \centering
\includegraphics[width=7cm]{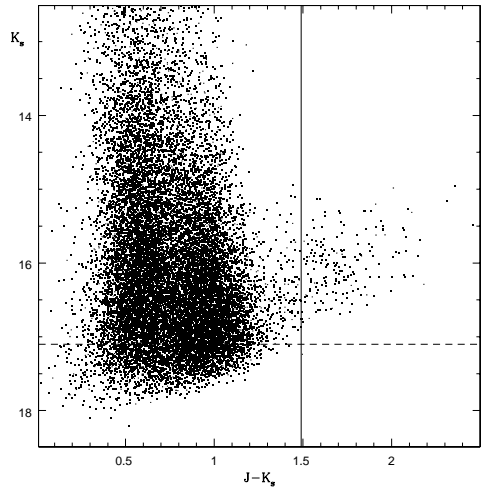}

   \caption{Colour-magnitude diagram of the combined NE and SW fields.
 The line vertical marks the blue limit
of the C stars region while the horizontal line indicates the magnitude of  
the TRGB.}               
              \label{FigCMD}
    \end{figure*}

\subsection{Distinction between stellar and non-stellar objects}

We see, in Fig. 2,  192 objects with magnitudes and colours corresponding to
AGB C stars. They have $(J-K_s)$ $>$ 1.49 and are well above the TRGB. 
This represents an unexpectedly large number,
when compared to the spheroid density profile,
 as determined by Battinelli et al.
(2006). Pollution by background galaxies has been noticed by Battinelli et al.
(2006) in the $(r' -  i')$ vs $(g' - r')$ colour-colour diagram of 
NGC 6822. It is thus possible that we see here interlopers which have magnitudes
and colours similar to C stars.

Even though
Galactic halo C stars can be well isolated with the SDSS colours
(Margon et al. 2002;
Downes et al. 2004), C stars at 500 kpc or more, 
with apparent magnitude $i'$ = 19 or fainter
can be confused with galaxies. 
Unfortunately, in  the $K_s$ vs $(J-K_s)$  colour-magnitude diagram, red galaxies can
also pollute the C star region of the CMD when the magnitude limit 
is deep enough.
The median $(J-K)$ colour of galaxies observed by Bershady et al. (1998), 
in the 17.5 $< K_s <$ 18.5 range, is 1.73. 
This colour is within our C star region but
the magnitudes are, however, fainter than the NGC 6822 C stars. 
Statistics from the first few fields of the MUNICS near infrared survey 
(Drory et al. 2001) indicate that
there should be $\sim$ 400 galaxies in our two CPAPIR fields within
 the magnitude interval: 15.25 $< K_s <$ 16.25, bracketing well the 
C star region. Not all these
galaxies are red enough, however, to be in the right colour interval. 
From the few
colour data they published,
 we calculate that 11/56, or 20\% of those galaxies are redder than
$(J-K)$ = 1.49, thus 80 $\pm$ 35 interlopers are expected among the candidate
``C stars'' of Fig. 2. These galaxies 
are bright enough to be in the not so distant Universe, most of them 
should therefore appear non stellar on our images. 

DAOPHOT-II provides an image quality diagnostics SHARP. For isolated
stars, SHARP should have a value close to zero, whereas for semi
resolved galaxies
and unrecognized blended doubles SHARP will be significantly greater than zero.
On the other end, bad pixels and cosmic rays produce SHARP less than zero. 
SHARP 
must be interpreted as a function of the apparent magnitude of all objects
because the SHARP parameter distribution flares up near the magnitude limit;
see Stetson \& Harris (1998) for a discussion of this parameter.
Since we want to investigate C star candidates, we select only stars
within the magnitude interval of C stars: 14.5 $<$ K$_s$ $<$ 16.75.
The SHARP parameter distribution of some 10,000 stars in this magnitude
range is 
presented in the upper panel of Figure 3. A Gaussian is fitted to the 
data. A slight surplus, above the Gaussian is seen for large positive 
SHARP values.
Contrary to the bulk of stars, the 192 C star candidates have a SHARP
distribution that confirms the presence of numerous 
non-stellar objects. 
We conclude from this figure that few genuine stars have SHARP larger
than $\approx$ 0.30 

   \begin{figure*}
   \centering
\includegraphics[width=7cm]{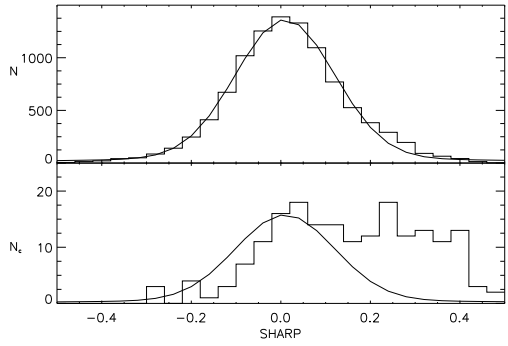}

   \caption{A comparison of the SHARP distributions for stars in the 
AGB magnitude interval. All stars are shown in the top panel and the
C star candidates in the lower panel.
The Gaussian, from above, is scaled down and reproduced below.
}
              \label{sharp histograms}
    \end{figure*}

For an alternative approach to the non-stellar pollution, 
we cross identify the CPAPIR database with the known C stars, identified  
from their $(R - I)$ and $(CN - TiO)$ colour indices, by Letarte et al. (2002). 
Our two CPAPIR fields
overlap a little with the CFH12k area. Sixty seven
 matches within 1.3$''$ are found.
The comparison of the SHARP parameter distributions for the C star candidates,
displayed in the lower panel of Fig. 3 is compared in Figure 4 with the 
the SHARP distribution of genuine C stars. 
Fig. 4 reveals that 95\% of the C stars
have a SHARP $<$ 0.30, we adopt this upper limit for the definition of
a stellar object. One must note that the pixel size of the CFH12k camera
is 0.206$''$, five times smaller than the CPAPIR pixel.

   \begin{figure*}
   \centering
\includegraphics[width=7cm]{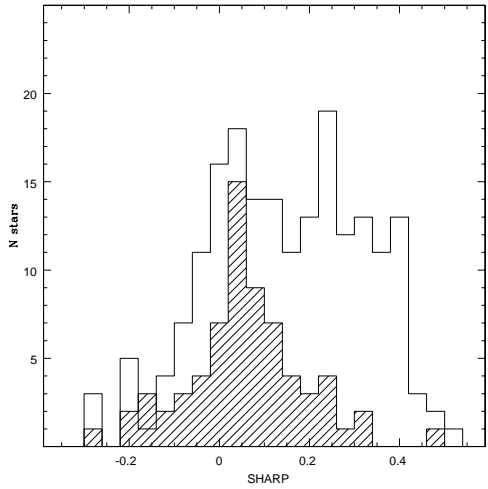}

   \caption{SHARP distribution of the 192 candidate C stars compared to
and the 67 matches to genuine C stars (shaded area). 
Barely 5\% of the genuine C stars
have a SHARP $>$ 0.30.
}           
              \label{SHARP}
    \end{figure*}

\subsection{The carbon stars}

There are then 142 stars with SHARP $<$ 0.30 and within the colour and
magnitude range of C stars.  If the rejected candidates are indeed galaxies,
their number is within the uncertainty of the expected number of galaxies
quoted in the previous section. The J2000.0 equatorial 
coordinates, magnitude and colour  of the C stars
are listed in Table 2. The numbering starts at 1001 in order not to
confuse them with Letarte et al. (2002) list of some 900 C stars previously
discovered in NGC 6822. Their  SHARP, as determined by DAOPHOT
 are also given in the last
column. These stars are not all new discoveries
 because of the mentioned  overlap between our fields
and the CFH12k field. The 68 previously known C stars (Letarte et al. 2002)
are marked with  a C while those found to be M stars, as explained in the next
section, are marked with a M. 
The Letarte et al. number is given in column 2, 000 is entered when no
match is found.
We calculate $\langle K_s\rangle$ = 16.02 $\pm$ 0.41, to be compared to 
   $\langle K_s\rangle$  = 15.82  $\pm$ 0.49 for the sample of Kang et al. (2006) and 
$\langle K_s\rangle$ = 15.69  $\pm$ 0.41 for the sample of Cioni \& Habing (2005).
$\langle (J-K)\rangle$ are respectively, 1.74, 1.90 and 1.81. 
 We have 21 C stars in common with the Cioni \& 
Habing (2005) list, the differences in mean magnitudes and mean colours are 
within the scatter.  
The comparison of their $K_s$ magnitudes, taken at different epochs,
 fails to reveal a substantial population of
variables among these C stars. For only two stars the $K_s$ differ by more than 0.4 mag.
One of them, with a $\Delta K_s$ = 0.47 mag, is a known C star for which we have 
its $I$ magnitude which can be compared with the measure of Cioni \& Habing (2005).  
Its $\Delta I$ = 0.03 mag, certainly not suggestive of a variation. For the second star,
the $\Delta K_s$ amounts to 1.3 mag, a huge value for a Mira. However, its $\Delta I$ 
= 0.06 again ruling out substantial variation. We currently have an ongoing Mira
survey in NGC 6822, within a year we should have sufficient observations to
identify red variables.

  \begin{table}
      \caption[]{C stars in the outer spheroid of NGC 6822{$^{\mathrm{a}}$}}
    $$
       \begin{array}{ccccccccr}
            \hline
            \noalign{\smallskip}
            id&id2&RA&Dec&K_s&\sigma_{K_s}&(J-K_s)&\sigma_{(J-K_s)}&SHARP\\
           \noalign{\smallskip}
            \hline
            \noalign{\smallskip}

  1001&000&19:42:38.66& -14:50:40.70& 16.379& 0.054& 1.782& 0.089& 0.03\\
  1002&000&19:42:43.66& -14:54:35.00& 16.615& 0.049& 1.666& 0.084&-0.05\\
  1003&000&19:42:45.52& -14:53:54.20& 16.540& 0.059& 1.720& 0.093& 0.17\\
  1004&000&19:42:48.95& -15:01:15.40& 16.191& 0.048& 2.076& 0.097&-0.02\\
  1005&000&19:42:50.17& -15:03:22.00& 16.427& 0.068& 1.636& 0.087& 0.22\\
  1006&000&19:42:52.92& -15:07:32.00& 16.229& 0.047& 1.543& 0.081&-0.09\\
  1007&000&19:43:02.72& -14:45:56.00& 16.377& 0.060& 1.994& 0.106& 0.14\\
  1008&000&19:43:04.89& -14:47:35.70& 16.450& 0.066& 1.900& 0.119& 0.29\\
  1009&000&19:43:13.39& -14:42:20.00& 16.541& 0.069& 1.614& 0.101& 0.29\\
  1010&000&19:43:17.65& -14:59:26.10& 16.321& 0.054& 1.849& 0.077&-0.03\\
  1011&000&19:43:17.99& -14:58:29.80& 16.290& 0.057& 2.182& 0.098& 0.20\\
  1012&000&19:43:18.60& -14:54:02.10& 16.275& 0.051& 1.706& 0.075& 0.02\\

            \noalign{\smallskip}
            \hline
         \end{array}
     $$
\begin{list}{}{}
\item[$^{\mathrm{a}}$] Table 2 is presented in its entirety in the electronic 
edition of Astronomy \& Astrophysics.
A portion is shown here for guidance regarding its
form and content. Units of right ascensions are hours, minutes and seconds, and
units of declination are degrees, arcminutes and arcseconds.

\end{list}
   \end{table}
\section{Discussion}

The identification of C stars from their NIR colours or from their 
$(CN-TiO)$ index are not entirely equivalent, as Demers et al. (2002) have
demonstrated. Our adopted $(J - K_s)$ limit for C stars is not expected to yield
a sample identical to the one obtained from other bands. In other words,
some of our newly identified C stars might be outside the box of C stars in
 the $(R-I$ vs $(CN-TiO)$ plane.
 To verify this possibility,  we cross identify the new C stars with our full CFH12k
database. Eighty five matches are found. These stars are plotted on the
colour-colour diagram of Figure 5. Sixteen stars must be classified M stars
on the basis of their $(CN-TiO)$ index. It turns out that the $(J - K_s)$ colours of these stars are
close to the blue limit, with a $\langle (J-K_s)\rangle$ = 1.57 while the 
142 C star candidates have $\langle J-K_s\rangle$ = 1.75. A few more are
marginal C stars very near the limit of acceptance.

   \begin{figure*}
   \centering
\includegraphics[width=6cm]{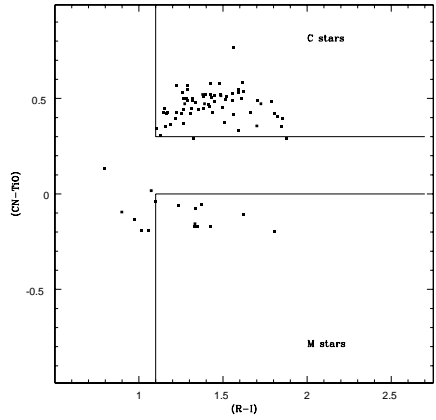}

   \caption{Colour-colour diagram of C stars candidates
matched to CFH12k database. Most stars with $(J-K_s)$ colours close to the
blue limit are actually M stars.
}
              \label{Fig cc}
    \end{figure*}

Likewise, if we consider the M star region of the CMD
we should expect to find some C stars among the M stars. Figure 6 presents the
$(R-I)$ vs $(CN-TiO)$ of stars with observed $(J - K_s)$ between 1.20 and
1.49 and matched to our CFH12k database.This is the region where M stars 
are found, see Cioni et al. (2004) for details. 
Of the 207 stars plotted less than 10\% have $(CN-TiO)$ corresponding to
C stars. With a $\langle I\rangle$ = 19.24 these 19 C stars have normal
luminosity in I, their $\langle K_s\rangle$ = 16.21 is, as expected, fainter
than the $\langle K_s\rangle$ of the redder C stars.

   \begin{figure*}
   \centering
\includegraphics[width=6cm]{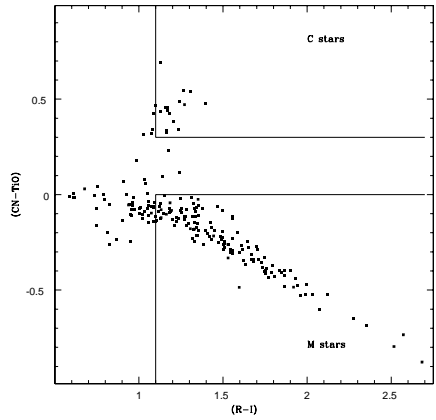}

   \caption{Colour-colour diagram of M stars, selected from their $(J - K_s)$
and matched to CFH12k database. Less than 10\% are actually C stars.
}
              \label{Fig cc for M}
    \end{figure*}

\subsection{SDSS colours of C stars}

We can, using our MegaCam data (Battinelli et al. 2006) cross identify 
the C star candidates and obtain their  $i'$, $(r'-i')$
and $(g' - r')$.  We recover magnitude and colours for 136 of the 142 C stars, they are listed
in Table 3, where we give $i'$, $(r'-i')$ and $(g'-r')$ along with 
their respective errors. Missing colours are entered as 9.999.
We presume that the few missing stars fall in the gaps of the MegaCam mosaic. 
A few stars, not necessarily the faintest do not
have $g'$ magnitude. The stars of Table 3 are plotted on the colour-colour diagram presented in  
Figure 7. The parallel lines define approximately the C star region of NGC 6822, as determined
by Battinelli et al. (2006) from the dereddened colours of known C stars. 
  \begin{table}
      \caption[]{SDSS magnitude and colours of the C stars{$^{\mathrm{a}}$}}
    $$
       \begin{array}{lcccccc}
            \hline
            \noalign{\smallskip}
            id&i'&\sigma_{i}&(r'-i')&\sigma_{r-i}&(g'-r')&\sigma_{g-r}\\
           \noalign{\smallskip}
            \hline
            \noalign{\smallskip}

   1001& 20.202&  0.025&  0.779&  0.034&  2.299&  0.057\\
   1002& 20.322&  0.024&  0.682&  0.031&  1.803&  0.039\\
   1003& 20.163&  0.022&  0.624&  0.029&  1.271&  0.029\\
   1004& 20.872&  0.029&  0.812&  0.040&  2.226&  0.068\\
   1005& 19.860&  0.018&  0.730&  0.024&  1.979&  0.035\\
   1006& 19.942&  0.014&  0.913&  0.022&  2.590&  0.048\\
   1007& 20.017&  0.019&  0.559&  0.023&  0.794&  0.019\\
   1008& 20.045&  0.018&  0.529&  0.022&  1.149&  0.020\\
   1009& 19.243&  0.011&  0.460&  0.014&  0.896&  0.014\\
   1010& 21.125&  0.043&  1.050&  0.066&  1.446&  0.080\\
   1011& 22.161&  0.089&  1.277&  0.151&  9.999&  9.999\\
   1012& 21.921&  0.055&  0.510&  0.067&  2.433&  0.104\\

            \noalign{\smallskip}
            \hline
         \end{array}
     $$
\begin{list}{}{}
\item[$^{\mathrm{a}}$] Table 3 is presented in its entirety in the electronic
edition of Astronomy \& Astrophysics.
A portion is shown here for guidance regarding its
form and content. 
\end{list}
   \end{table}

   \begin{figure*}
   \centering
\includegraphics[width=6cm]{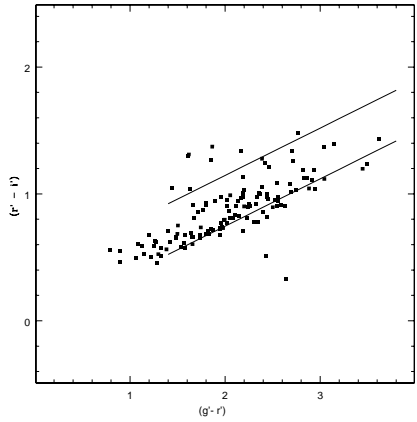}

   \caption{Colour-colour diagram of C stars seen in the MegaCam database. The 
parallel lines limits the C star region, as defined from the unreddened 
colours, for NGC 6822 by Battinelli et al. (2006).
}
              \label{Fig SDSS colour-colour}
    \end{figure*}

In Figure 8 we compare the i$'$ magnitude distribution of the known C stars
with the newly discovered ones, shown by the shaded
histogram. The $\langle i'\rangle$ of the two sets are respectively 19.90 and
20.03. This essentially confirms that we are dealing with genuine C stars.

   \begin{figure*}
   \centering
\includegraphics[width=6cm]{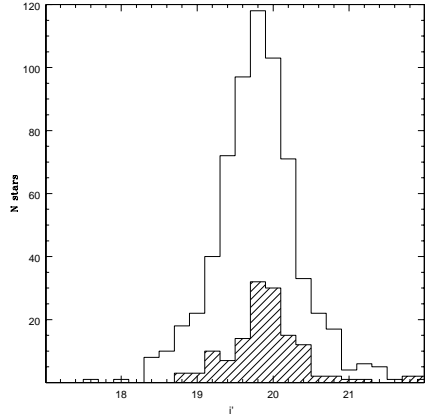}

   \caption{Magnitude distributions of known C stars and of the newly discovered
ones (shaded area). Both sets have essentially the same distribution.
}
              \label{SDSS histograms}
    \end{figure*}

\subsection{Spatial distribution of AGB stars}
Battinelli et al. (2006) have shown that C stars, as defined from their 
SDSS colours, can be traced in the spheroid of NGC 6822 to $\sim$ 40$'$.
C stars, selected from their $(J-K_s)$ colours, could be used to confirm this
finding. Contrary to the MegaCam observations, the near infrared
 observations have not
surveyed the whole spheroid.  
Cioni \& Habing (2005) observed  the central $20'\times 20'$ area
of the galaxy while our two fields are aligned along the outer major axis.
Nevertheless, a band of $\pm 7'$ along the major axis of the spheroid,
(at position angle of 65${^\circ}$) is fully sampled thus it could be used to
determine the C star density profile along the major axis.
We display, in Figure 9, the surface density (stars per arcmin$^2$) 
profile for C stars along the major axis. The solid curve, defined from 
10$'$ to 50$'$, corresponds to the two-exponential profile determined from RGB 
star counts by
Battinelli et al. (2006). It has been shifted down to match the points.
Even if our figure is plagued by small number statistics,
the outer bins contain just a few stars each, the C stars appear to 
follow the density of the bulk of the spheroid.

   \begin{figure*}
   \centering
\includegraphics[width=6cm]{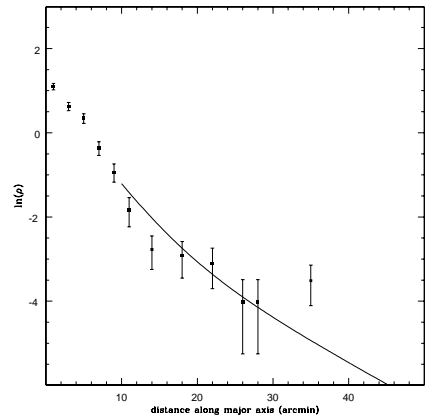}

   \caption{Surface density profile of C stars along the major axis of
the spheroid of NGC 6822. The solid line represents the double exponential
profile established from RGB stars.
}
              \label{profile}
    \end{figure*}

\section{Conclusion}
The main goal of the present paper 
was to find bright kinematic probes, namely C stars, to trace the rotation 
curve of the NGC 6822 spheroid up to its detected limit ($\approx 36'$).
We have obtained accurate $J$,$K_s$ photometry for stars in two 34$'\times 34'$ fields located
in the periphery of the  spheroid along the major axis. On the basis of 
their $(J- K_s)$ 
colours we find 142 {\it bona fide} C stars, half of them are 
newly discovered objects. 
Maps of these C stars are shown in Figure 10, for the 
NE field, and in Figure 11 for the SW field. Solid dots represent known 
C stars while the
new ones are indicated by open circles for those with SHARP $<$ 0.20, (75\% 
of them) and open
triangles for those with 0.20 $<$ SHARP $<$ 0.30. As expected a few of the
high SHARP objects match known C stars. However, one notes a unmistakable 
surplus of triangles in the areas corresponding to the outer spheroid. 
It is expected that a number of high SHARP candidates are non-stellar.
Nevertheless, future observers can prioritize their candidates from the
SHARP given.
The inner square traces the border of the
CPAPIR field. These new C stars will permit to more than double the length of the
rotation curve of the spheroid and reach even further than the HI survey.

   \begin{figure*}
   \centering
\includegraphics[width=6cm]{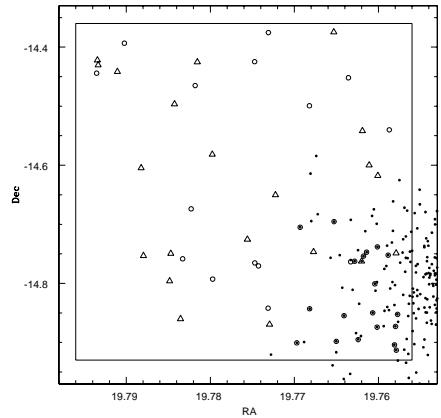}

   \caption{Map of the NE CPAPIR field, C stars found in the CFH12k survey
are shown by solid dots while the C stars 
identified here are represented by circles, for those with SHARP $<$ 0.20 and
open triangles for stars with 0.20 $<$ SHARP $<$ 0.30
}
              \label{NE field}
    \end{figure*}
 
   \begin{figure*}
   \centering
\includegraphics[width=6cm]{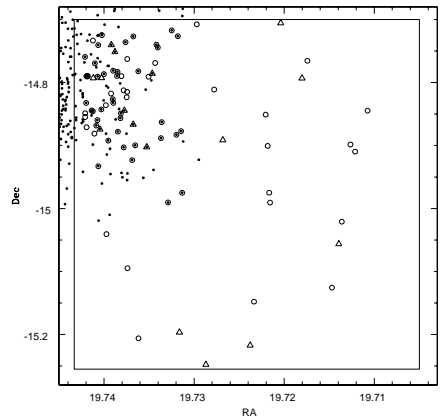}

   \caption{Map of the SW CPAPIR field, C stars found in the CFH12k survey
are shown by solid dots while the C stars 
identified here are represented by circles, for those with SHARP $<$ 0.20 and
open triangles for stars with 0.20 $<$ SHARP $<$ 0.30
}
              \label{SW field}
    \end{figure*}

\begin{acknowledgements}
This research
is funded in parts (S. D.) by the Natural Sciences and Engineering Research
Council of Canada. 
\end{acknowledgements}

\end{document}